\documentclass[10pt]{iopart}

\usepackage{graphicx}

\usepackage{hyperref} % For hyperlinks in the PDF

\newcommand{\pr}{\partial}
\newcommand{\rta}{\rightarrow}

\newcommand{\ep}{\epsilon}

\newcommand{\p}{\prime}
\newcommand{\om}{\omega}
\newcommand{\ra}{\rangle}

\newcommand{\beq}{\begin{equation}}
\newcommand{\eeq}{\end{equation}}

\newcommand{\ben}{\begin{enumerate}}
\newcommand{\een}{\end{enumerate}}

\begin{document}

\title[Unconventional superconductivity in cuprates]{Is the problem of Cuprate high-$T_c$ superconductivity a solved problem?}

\author{Navinder Singh}

\address{Theoretical Physics Division, Physical Research Laboratory, Ahmedabad. India. PIN: 380009.}
\ead{navinder.phy@gmail.com; navinder@prl.res.in}
\vspace{10pt}
\begin{indented}
\item[]19th Oct 2022
\end{indented}

\begin{abstract}
The recent experimental verification of the charge-transfer superexchange mechanism as the microscopic pairing mechanism of high-$T_c$ cuprate superconductivity by Seamus Davis and collaborators\cite{sea} is a tour de force! The correct model for cuprates is the three band Emery model in which oxygen p-orbitals are explicitly taken into account. The doped holes go into these oxygen p-orbitals where they undergo charge-transfer superexchange with unpaired electrons in copper d orbitals. This charge transfer superexchange is the key which leads to bound pairs and superconductivity. In the experimental verification\cite{sea}, the system chosen is $Bi_2Sr_2CaCu_2O_{8+x}$. What is achieved is the direct functional dependence of the local electron pair density ($n_P(r)$) on local charge transfer energy ($\ep_{pd}(r)$) using state of the art single-electron and electron-pair (Josephson)  scanning tunneling microscopy. The quantitative functional dependence of $n_P(r)$ on $\ep_{pd}(r)$ matches with that indicated and deduced by theory\cite{t1,t2,t3,t4,t5,t6,t7,t8,t9,t10}. The verdict of the experiment settles the debates on the microscopic mechanisms of the cuprate superconductivity in the clear favor of charge-transfer superexchange mechanism. We discuss this development in brief, and present a simple minded approach to the essence of cuprate superconductivity. We discuss what is settled now, and what is not settled yet. A "theoretical minimum" of the high-$T_c$ problem is also discussed.
\end{abstract}

%
% Uncomment for keywords
\vspace{2pc}
\noindent{\it Keywords}: Unconventional superconductivity in Cuprates, Charge-transfer superexchange mechanism, Pseudogap.
% Uncomment for Submitted to journal title message
%\submitto{\JPA}
%
% Uncomment if a separate title page is required
%\maketitle

\vspace{2pc}

\noindent{``...Thus when one has found {\it A} way through the maze of conflicting requirements, that is certain to be {\it THE} way, no matter how many deep-seated prejudices it may violate..." ---P. W. Anderson.}

\vspace{2pc}

% 
% For two-column output uncomment the next line and choose [10pt] rather than [12pt] in the \documentclass declaration
\ioptwocol

\section{Recent experimental verification of charge-transfer superexchange in brief}
%%%%%%%%%%%%%%%%%%%%%%
There is a growing consensus in the community that the single band Hubbard model cannot fully address the physics of the hole doped cuprates, even though it has some success\cite{t9,t10,shen}. In addition, the direct experimental evidence has shown that hole doping removes electrons from oxygen p-orbitals\cite{gau}. Thus a three band model that incorporates oxygen p-orbitals explicitly is a must to fully address the key to the pairing mechanism.  Very early on this job was completed by V. J. Emery\cite{emery}. And he rightly pointed out that key to the mechanism is the exchange interaction (exchange interaction between un-paired electrons on oxygen p-orbitals (created due to hole doping) and an un-paired electrons in the central Cu d$_{x^2-y^2}$-orbitals). He could predict right order of magnitude of $T_c$ with his three band model, now called the Emery model.  P. W. Anderson also rightly guessed that the key to the pairing mechanism is the exchange interaction (superexchange) not the conventional phonon mediated interactions\cite{ander1,ander2}. But he insisted on the single-band Hubbard model as a low energy model appropriate for cuprates\cite{ander2}. A reduction from the Emery model to a single band Hubbard model was also achieved\cite{zr}.  But now we understand that oxygen p orbitals must be taken into account explicitly as charge-transfer energy (from p-band to d-band) is the key parameter that controls pairing\footnote{Although the Zhang-Rice reduction from three-band to single-band model incorporates the charge transfer energy ($\ep_{pd}$) in $J=\frac{4 t_0^4}{\ep_{pd}^2 U} +\frac{4t_0^4}{2 \ep_{pd}^3}$, but the single band $t-t^\p-t^{\p\p}-J$ model proved to be insufficient with respect to other probes\cite{shen}.}. The experiment by Seamus Davis and collaborators has unequivocally settled it!

Before we present a simple model that explains the key to the pairing mechanism in cuprates superconductors in layman language, we sketch a brief review of this crucial experiment\cite{sea}.

Authors\cite{sea} achieved a coterminous visualization of the local charge transfer energy $\ep_{pd}(r)$ and local electron pair density $n_P(r)$ by measuring the dependence of these quantities on $\delta(r)$, which is the distance of the apical oxygen atom from the planner copper atom (just below it).  They selected the canonical cuprate $Bi_2Sr_2CaCu_2O_{8+x}$ at optimal doping $p=0.17$ in which there is a crystal supermodulation in $\delta(r)$ with a period of 26\AA. They measure the differential conductance $g(r,V)$ as a function of the location $r$ and the tip-sample voltage $V$. From which they derive the spatially resolved EDOS ($N(r,E) \propto g(r, V=E/e)$) and from that the local charge transfer energy $\ep_{pd}(r)$ is derived as the minimum distance between upper and lower bands at constant conductance\cite{sea}. Thus a $\delta(r)$ dependent $\ep_{pd}(r)$ is determined.

Then using the superconducting scanning tunneling microscope (STM) tips (of the same material), Josephson critical current (for electron pair tunneling) is also measured as a function of location $r$ (thus $\delta(r)$) from which electron pair density $n_P(r)$ is determined as a function of $r$ (thus $\delta(r)$). 

They find that the charge transfer energy decreases linearly as $\delta$ increases and the pair density increases linearly as $\delta$ increases. By eliminating the common variable $\delta$ they find that the electron pair density decreases linearly as the charge transfer energy increases ($\frac{dn_P}{d\ep} \simeq -0.81 \pm 0.17~eV^{-1}$). This value agrees reasonable well with theory ($\frac{dn_P}{d\ep} \simeq -0.9~eV^{-1}$)\cite{t9}.   This is intuitively understandable as the pairing energy scale $J$ roughly varies at $t^4/\ep_{pd}^3$, that is, increase in the charge transfer energy reduces the pair forming superexchange energy $J$.

It is a clear proof that the charge transfer superexchange is the key which leads to bound pairs. The experimental and theoretical agreement of these quantities settles the debates on the microscopic mechanisms of the cuprate superconductivity in a clear favor of the charge-transfer superexchange mechanism.

In the next section we explain the charge transfer exchange with a very simple model to illustrate the pairing mechanism.

\section{The charge-transfer exchange mechanism in a simple language}
%%%%%%%%%%%%%%%%%%%%%%%%%%%%%%
To understand the essence of the mechanism behind cuprate superconductivity in a simple language, consider figure (1). There, four states are depicted $|1\ra, ~|2\ra,~|3\ra,~$ and $|4\ra$. There are un-paired electrons in copper $3d_{x^2-y^2}$ orbitals with opposite spins, and there is an un-paired electron on oxygen $p_\sigma-$orbital (after hole doping, as hole is created in oxygen $p_\sigma$ orbital, leaving an un-paired electron there). In state $|1\ra$ the spin in oxygen $p_\sigma$ orbital is polarized in the $+z$ direction and in state $|2\ra$ the spin is polarized in the $-z$ direction. Now consider that the un-paired electron on the copper ion on the left hand side tunnels into oxygen $p_\sigma$ orbital making it doubly occupied (state $|3\ra$). It costs energy, the charge transfer energy and Coulomb repulsion between two electrons in the oxygen $p_\sigma$ orbital. We denote the total energy cost with $\ep_{pd}$. The tunneling happens with an amplitude $-t_{pd}$. There is another possibility that the un-paired electron in the oxygen $p_{\sigma}$ orbital tunnels into copper $3d_{x^2-y^2}$ orbital, making it doubly occupied. But that costs a large amount of energy $U$ and that configuration we do not consider (No double occupancy in $3d_{x^2-y^2}$ orbital of copper!). In state $|4\ra$  the un-paired electron from the copper ion on right hand side tunnels into oxygen $p_\sigma$ orbital making it doubly occupied. 

\begin{figure}[!h]
\begin{center}
\includegraphics[height=6cm]{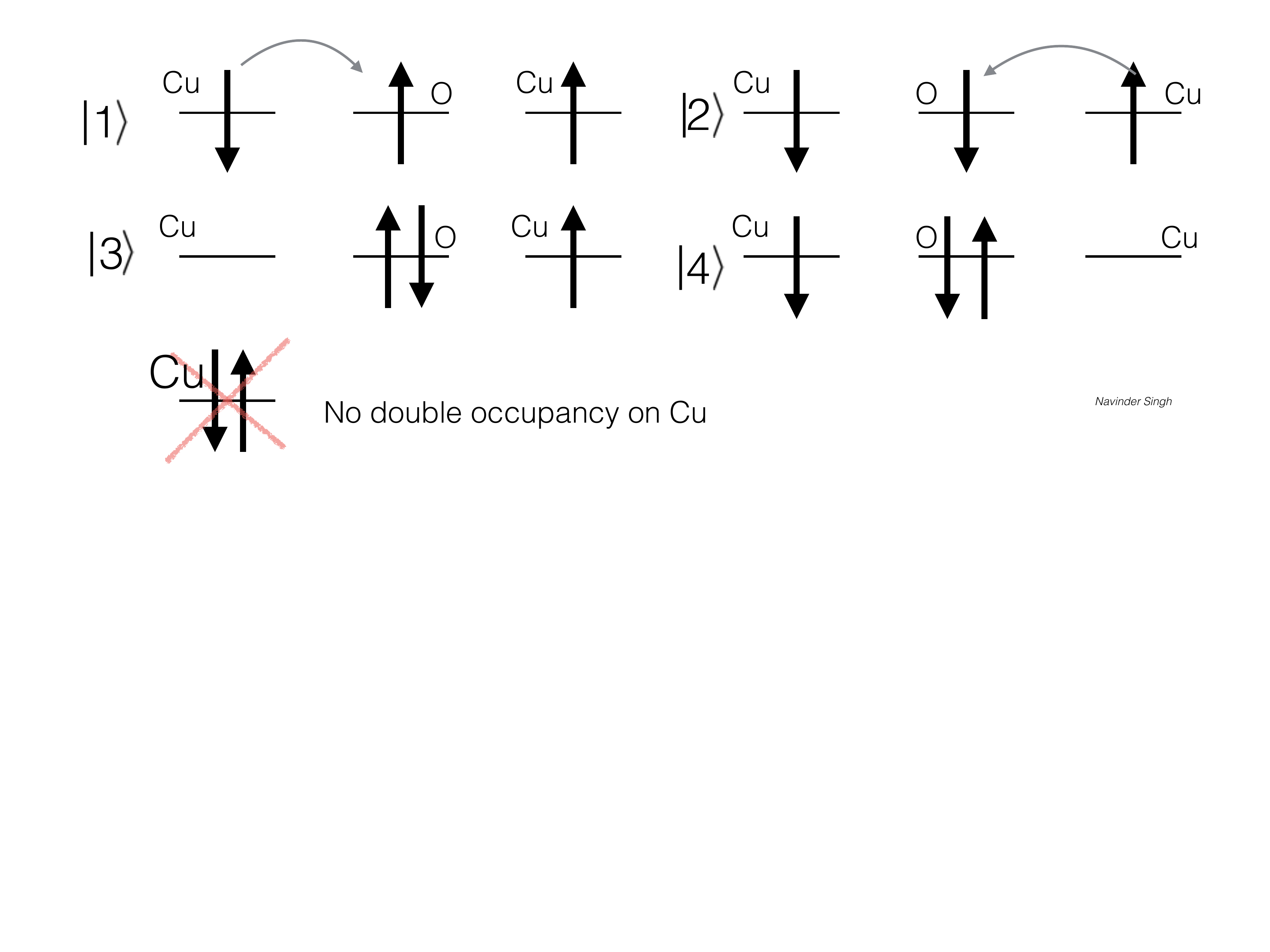}
\caption{Simplest exchange mechanism between an un-paired electron on copper atom and an un-paired electron (after hole doping) on oxygen atom.}
\end{center}
\end{figure}

These states are not still or stationary in time. The electron tunnels back and forth between copper $3d_{x^2-y^2}$ orbital and oxygen $p_\sigma$ orbitals. These "flip-flops" happen on a time scale of femto-seconds ($\sim\frac{\hbar}{\ep_{pd}}$). This exchange leads to a lower energy state. The solution is simple. Our Hilbert space is 4 dimensional with basis states $|1\ra, ~|2\ra,~|3\ra,~$ and $|4\ra$. We want to find the eigen-energies.  We will write the equations of motion in the site basis and transform them into energy basis to see this. The system will be in the superposition state : $|\chi\ra = C_1 |1\ra +C_2|2\ra+C_3|3\ra+C_4|4\ra$, and the coefficients follow the EOM:

\beq
i\hbar \dot{C}_i = \sum_j H_{ij}C_j.
\eeq

We set the energy (in site basis) of the states $|1\ra$ and $|2\ra$ as $\ep_0 =0$ and that of $|3\ra$ and $|4\ra$ as $\ep_0 + \ep_{pd} =\ep_{pd}$. The non-zero matrix elements of the Hamiltonian are $H_{13}=H_{31}=H_{24} =H_{42} = -t_{pd}$ and $H_{33}=H_{44} = \ep_{pd}$, and rest of all are zero. It is easy to show that we end up having coupled (but in pairs) linear differential equations:

\begin{eqnarray}
i\hbar \dot{C}_1 &=& -t_{pd}C_3\\
i\hbar \dot{C_3} &=& -t_{pd}C_1 + \ep_{pd} C_3\\
i\hbar \dot{C}_2 &=& -t_{pd}C_4\\
i\hbar \dot{C_4} &=& -t_{pd}C_2 + \ep_{pd} C_4
\end{eqnarray}

The solution is easy to obtain with the following eigenvalues of energy:

\beq
E_{\pm} = \frac{1}{2}(\ep_{pd} \pm \sqrt{\ep_{pd}^2 + 4 t_{pd}^2})
\eeq

Thus, there exists a $-ve$ energy solution (a bound state) with energy $E_{-} =\frac{1}{2}(\ep_{pd} - \sqrt{\ep_{pd}^2 + 4 t_{pd}^2})$, which is $~\sim -\frac{t_{pd}^2}{\ep_{pd}}$ when $t_{pd}\ll\ep_{pd}$. If the tunneling is stopped ($t_{pd} =0$) we get back unbound (positive and zero energy states) states, as we should.

{\bf Thus, it is the "flip-flops" (as Feynman put it\cite{feyn} in the context of ammonia molecule) of an un-paired electron between the oxygen p sigma orbital and the copper $3d_{x^2-y^2}$ orbital, which is the experimental, the theoretical, and the empirical crux of the matter! Pauling called it resonance lowering of energy\cite{pauling}. Kramers-Anderson called it exchange lowering (if the "flip-flop" happens between copper spins via oxygen orbitals, it is called the superexchange). V. J. Emery called it "exchange of holes"\cite{emery}. They meant the same mechanism. Their intuition has withstood the test of time! Once there are many bound pairs they can condense into the superconducting state below some critical temperature. }

\subsection{Single hole}
Consider a single hole doped into p sigma orbital of an oxygen atom (in what follows we discuss in terms of the hole language as is more appropriate for hole doped cuprates).  In a femto-second time scale ($\sim \hbar/\ep_{pd}$) it starts to resonate with the neighboring copper spins. Since all the oxygen sites around the given copper atom are equivalent, it starts to resonate via all the four oxygen p sigma orbitals. It lowers its energy further by doing this. But it does not stay there. It can still lower its energy by tunneling to a neighboring  plaquette of four oxygen atoms around one copper atom in a time scale of $\hbar/t$ where $t$ is the tunneling matrix element between two nearby plaquettes. In fact it will do a random motion in the entire $CuO_2$ lattice, each time, on its way, resonating with the un-paired copper spins (figure 2).  A certain fraction of them (at typically 5-6 percent hole doping) are effective in destroying the entire AFM spin alignment of copper spins. It is something to do with its quantum random motion within the entire copper oxide lattice.

\begin{figure}[!h]
\begin{center}
\includegraphics[height=6cm]{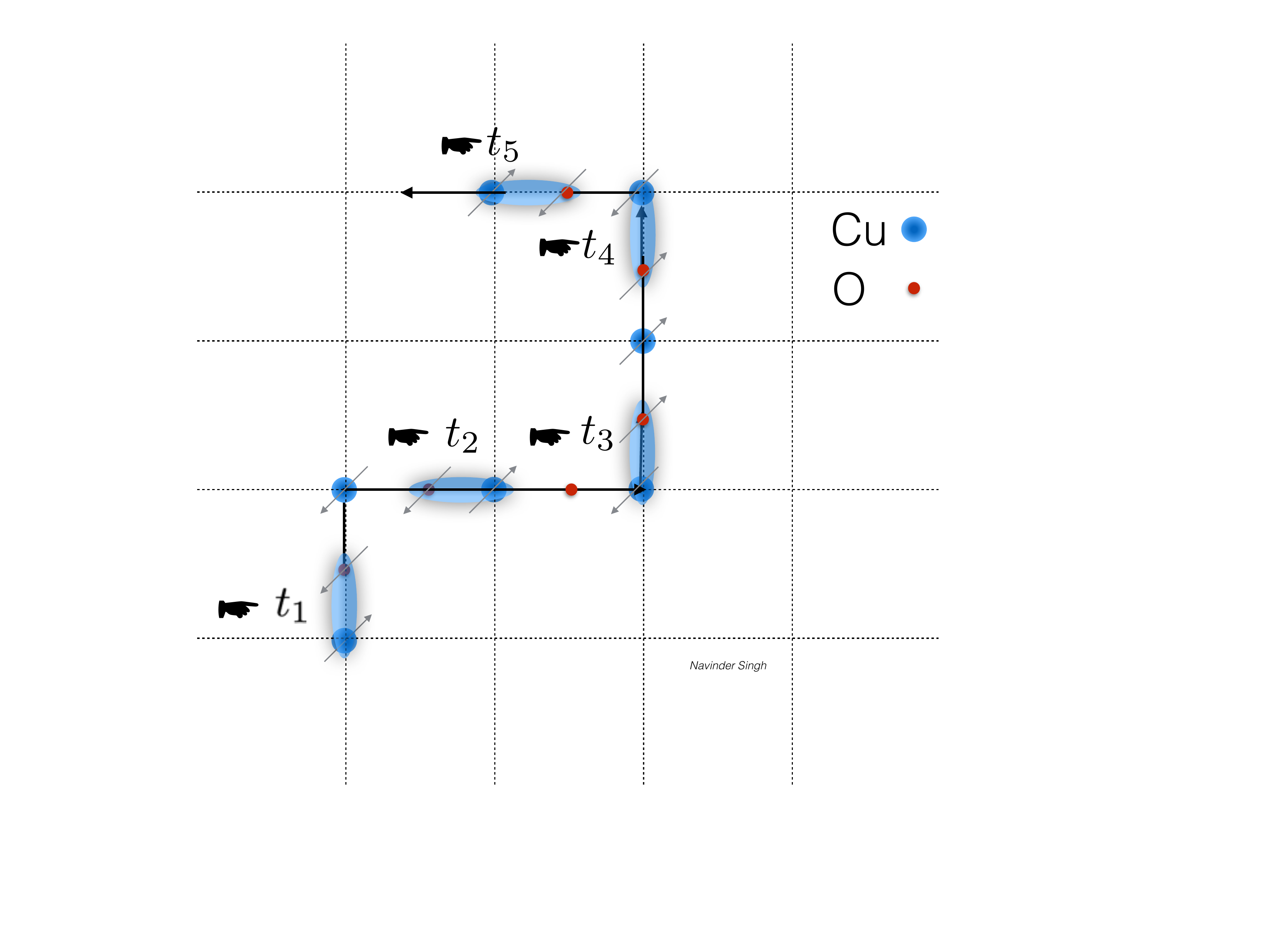}
\caption{Quantum random motion of a single doped hole, and the resonances it creates on its way. Picture depicts schematic snapshots at different times at $t_1,~t_2,~t_3,~t_4$, and at $t_5$.}
\end{center}
\end{figure}

\subsection{Two holes}
It is not difficult to imagine the possibility of further lowering of energy of the resonating holes in view the gain in $t$ terms in $E_{-} \sim -(\frac{t_{pd}^2}{\ep_{pd}} ~+~ extra~t~terms)$ when the two holes are around one copper spin. A further resonating process will set up: resonance of the hole on the copper with two nearby holes in oxygen p sigma orbitals. Mathematics becomes more complicated (but the essence of the mechanism remains the same). Thus two holes doped at two different locations will do random motion through the copper oxide plane, and ultimately they will "catch" each other and a new resonance process will set up which will further lower its energy. Then this hybrid (two holes around one cooper spin) will do random motion (as a single entity) in the entire lattice, because by doing this, it gains $t$ terms and further lowers its energy. Mathematical formulation of these physical pictures can be made but that will be of no consequence as experimental injection of a few holes in a copper-oxide layer and their experimental study is out of question. Also actual lattice is never perfect. The "two-hole-composite" may end up trapped around some lattice impurity killing its own random motion.

\subsection{N holes}
The topic of N holes brings us to Anderson's resonating valence bond picture. A complete mathematical treatment has been impossible so far. But then the problem is captured from another powerful intuitive argument: Superconducting state can be obtained from the mean-filed BCS wave-function of all the resonating holes provided double occupancy is restricted on any given copper spin. This and other alternative approaches implementing the Anderson's RVB idea are:\cite{ander1,ander2,ander3,param,zhang,ruck,kot,lee2,scal,bas,bas1}. An excellent review is presented in:\cite{lee}.

V. J. Emery solved this problem using a simple but powerful argument. He first showed that the exchange mechanism between un-paired electrons on oxygen atoms via an un-paired electron on the central copper atom (refer to figure 1 in his paper\cite{emery}) leads to an attractive interaction with energy $-v_0$ (expression of $v_0>0$ is given in his paper\cite{emery}).

In the second step he uses the simple BCS mean field theory and the BCS gap equation replacing phonon interaction by his exchange mechanism interaction. The estimated value of $T_c$ agrees reasonably well with experimental value. This also shows that a simple-minded  mean-field approach is not far from truth!

In what follows we present a perspective and some critical comments on the way field developed in the last 36 years.

%%%%%%%%%%%%%%%%%%%%%%%%%

\section{Anderson's "dogmas":}
%%%%%%%%%%%%%%%%%%%%%

Anderson's "dogmas" are some factual inferences obtained from a careful analysis of some of the important experiments and logical reasoning\cite{anderb}. The purpose of the "dogmas" is to get-rid-of the irrelevant information and to simplify the theoretical problem. Below we collect these dogmas; one by one:

\begin{center}
\begin{minipage}{.4\linewidth}
DOGMA NO. 1:  
\end{minipage}
\end{center}
Look at the copper oxide planes only: All the relevant carriers of spin and charge reside in the copper oxide planes, and are present in the anti-bonding band $Cu~3d_{x^2-y^2} -- O~2p_{\sigma}$ which is split into two Hubbard bands due to strong on-site interaction $U$.  All other bands are filled thus dead. Batlogg's and more recent observations of universal $\rho_{ab}$ (same in-plane resistivity per copper-oxide plane) is the direct experimental support to dogma no. 1\cite{anderb,bat,bari}.  This establishes the two dimensionality of the cuprate problem\cite{anderb}.

\begin{center}
\begin{minipage}{.4\linewidth}
DOGMA NO. 2:  
\end{minipage}
\end{center}

Magnetism and superconductivity: Magnetism and superconductivity are closely related. The very same electrons that order AFM at $x=0$ become charge carriers at finite doping, and below a critical temperature condense into copper pairs, and exhibit superconductivity. 

In defense, Anderson argues that at zero doping cuprates are Mott-Hubbard insulators and exhibit antiferromagnetism, and there are other compelling reasons from optical, photoelectron spectroscopy, and NMR data\cite{anderb}.

\begin{center}
\begin{minipage}{.4\linewidth}
DOGMA NO. 3:  
\end{minipage}
\end{center}

Dominant interactions are repulsive: Dominant interactions in copper-oxide planes are repulsive and their energy scales are large. The major role is played by $J$ and $U$. Other interactions like electron-phonon and electron-impurity are sub-dominant.

Basis of this "dogma" is that cuprates at zero doping are Mott-Hubbard insulators (not band insulators)\cite{anderb}.

\begin{center}
\begin{minipage}{.4\linewidth}
DOGMA NO. 4:  
\end{minipage}
\end{center}

Non-Fermi liquid: The electronic state ("normal" state) in copper-oxide planes is not a Fermi liquid in the sense that quasiparticle weight goes to zero ($Z\rta0$). It is the direct consequence of the experimentally inferred self-energy:
\beq
\Sigma^{''}\propto \omega,~~~~\Sigma^{'} \propto \omega \log(\omega),
\eeq
and  thus $Z = (1-\frac{\pr \Sigma}{\pr \omega})^{-1} \rta 0$. Therefore, the quasiparticles cannot be defined in the Landau sense. However, experiments show that Fermi surface exists, and obeys the Luttinger theorem.

Anderson advances serval arguments to defend this assertion\cite{anderb}. Leading ones are the anisotropy in c-axis and ab-plane transport, and several conclusions from ARPES and optical measurements\cite{anderb}.

Anderson's "dogma" 5 and 6 deal with his {\it interlayer hopping} theory, which has turn out to be wrong.

So we notice that the above mentioned "dogmas" of Anderson has withstood the test of time. Only change is in 'dogma no. 1'. The oxygen orbitals cannot be integrated out. Hole doping creates holes in oxygen $p_{\sigma}$ orbitals, and three band Emery model captures the essential physics.

Also we reproduced these "dogmas" here because these are essential facts derived from experiments, and any successful theory (be it for the superconducting state or for the strange metal regime) must respect these facts. The verdict of the experiment is in favor of the super-exchange based theories and these are in harmony with these dogmas as far as the superconducting state is concerned.
%%%%%%%%%%%%%%%%%%%%%

\section{What is settled now?}
%%%%%%%%%%%%%%%%%%%%%%%%%%%%%%%%%%%%%%%%%%%%%%%%%%%%%%

\begin{enumerate}

\item Pairing mechanism:

\begin{center}
\begin{minipage}{.4\linewidth}
Proposition No. 1:  
\end{minipage}
\end{center}

The Quantum Critical Point (QCP) which is the point (on T=0 line) where pseudogap ends (when superconductivity is killed in high magnetic fields) is the crux of matter in the sense that the fluctuations of the nearby antiferromagnetic order are the cause of pairing and superconductivity  and the strange metal regime is dominantly affected by it (as the scattering is strongest at the QCP, marked by several probes including a dead $T-$linear resistivity).

\begin{center}
\begin{minipage}{.4\linewidth}
Proposition No. 2:  
\end{minipage}
\end{center}

It is Emery's "exchange of holes" mechanism or Anderson's superexchange mechanism, and its RVB formulation which can explain pairing. But the strange metal regime is outside the scope of the current developments of this idea.

The question is which one of the above proposition is the correct one as far as the mechanism of superconductivity is concerned. This is a very difficult matter.  Author understands that the framing the issue in this way might be rather offensive! But, out of the two propositions, as far as the superconducting state is concerned, the proposition 2 has clearly gained weight with the experimental verification of the superexchange mechanism\cite{sea}. Thus, the author is in the defense of proposition 2. Exact connection of exchange physics (like RVB model) with QCP remained to be worked out, and more serious thought and discussions are required.

But as far as the pairing mechanism is concerned, proposition no. 2 must be defended!

\item Issue of phonons:

There is no isotope effect at the optimal doping and at this doping the DC resistivity is dead $T-$linear from very low mili-Kelvin temperatures to very high about thousand-Kelvin temperatures\cite{taillefer1,cooper}. There is no $T^5$ like bending at low temperatures at $T\ll T_D$(figure 3) which is a signature of phonon scattering. There are many other indications, and according to Anderson's dogma no. 3 phonons remain sub-dominant both in the superconducting state and in the strange metal regime. Other reasons that phonons play a sub-dominant role are discussed in\cite{ander3, anderb}.

\begin{figure}[!h]
\begin{center}
\includegraphics[height=3cm]{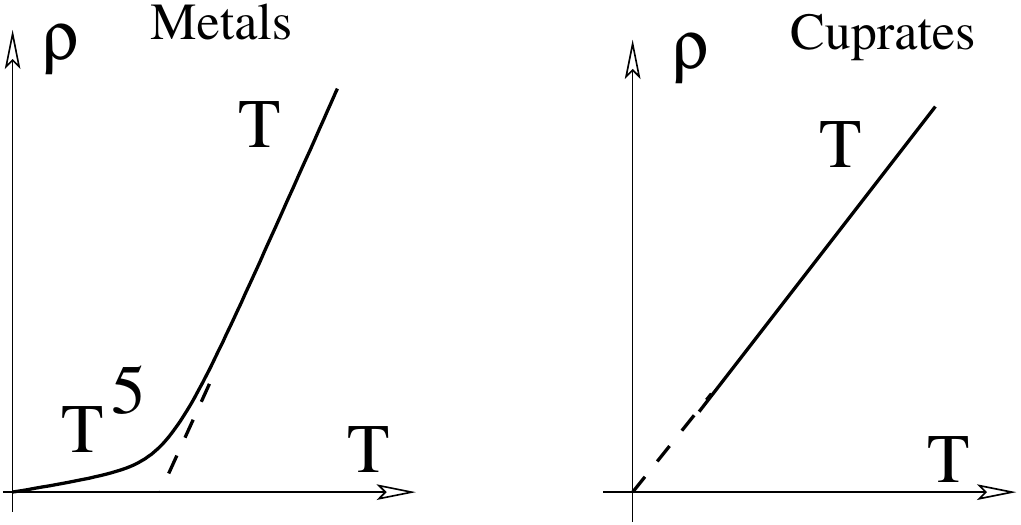}
\caption{DC resistivity in a simple metal and in cuprates.}
\end{center}
\end{figure}

\item 
The pairing is local in Cuprates! This is evident from the fact that cooper pair size is roughly of the order of $10~\AA$ to $30~\AA$\cite{dago}. This is from the indirect penetration depth measurements. Actual pair size may even be smaller. This is many orders of magnitude smaller than that found in the BCS superconductors (of the order of $1000~\AA$). Small size of the pairs does not mean that the mean-field approaches are not useful in the case of cuprates. V. J. Emery's solution using mean-field BCS gives a reasonable value of $T_c$!

%%%%%--------------------

\item Anderson's assertion that the idea of a pairing glue is a folklore has turned out to be true!  Both $U$ and $J$ leads to a very high energy (high frequency) dynamics and the Eliashberg type theories are not on the right track\cite{ander20}.

\end{enumerate}

\section{What is not settled yet?}
%%%%%%%%%%%%%%%%%%

\begin{enumerate}

\item Theoretical deduction of the $T-$linear resistivity in the strange metal regime is a major problem that remains.  More specifically, how to derive $A(x)\propto T_c(x)$? Where $A(x)$ is the coefficient in the $T-$linear term in the resistivity and $T_c(x)$ is the critical temperature at that doping\cite{taillefer1,taillefer2}.

\item Theoretical deduction of the non-Drude scattering rate $\frac{1}{\tau(\om)} \propto \om$ in a formalism which is in harmony with Anderson's superexchange physics is another open issue.

\item The rise of $T_c$ with doping (in the underdoped regime), that is the rising part of the "dome",  is understandable from the phase fluctuations and low superfuild density perspective (simply less number of cooper pairs at low doping)\cite{emery2}. But this author believes that the justice is not yet done with the mechanism why $T_c$ decreases with doping in the overdoped regime (that is fall of the "dome"). In the RVB model it is thought to be due to the decrease in the pairing scale due to the decrease in the renormalization factor $g_s(x)$ on doping\cite{ander3}. But as far as author understands they miss out a very crucial observation that the carrier number reduces from $1+p$ holes per $Cu$ atom at $p>p^*$ to $p$ holes per $Cu$ atom at $p<p^*$\cite{taillefer2}. The accurate quantitative theory that incorporates this very important fact is still lacking.

\item Some other open issues are discussed in\cite{nav}.

\end{enumerate}
%%%%%%%%%%%%%%%%%%%%%%%%%%%%%%%%%%%%%%%%%%%%%%%

\section{Whether "a theoretical minimum" of high-$T_c$ superconductivity possible?}
%%%%%%%%%%%%%%%%%%%%%%%%%%%%%%%%%%%%%%%%%%%%%%%%

The question of a "metric" or of a "theoretical minimum" of high-$T_c$ problem is vexed one.  Which experimental result is the key result is subjective to some extent. Let us try to design a set of sample "theoretical minima" and  then we will argue.

\begin{center}
\begin{minipage}{.4\linewidth}
Sample "theoretical minimum" no. 1:  
\end{minipage}
\end{center}

History has taught a lesson: the full understanding of the BCS theory came with the understanding of the normal state (the Fermi Liquid Theory).  FLT was in place when BCS arrived. Similarly, to have a complete understanding of the high-$T_c$ problem we must first understand the strange metal regime out of which (on cooling) the superconductivity emerges. For example, any successful theory should explain the marginal Fermi liquid phenomenology\cite{mflp}. Then the theory should explain right order of magnitude of $T_c$ and the "dome" shape structure with experimentally determined other required parameters.

\begin{center}
\begin{minipage}{.4\linewidth}
Sample "theoretical minimum" no. 2:  
\end{minipage}
\end{center}

The parent compounds are Mott-Hubbard insulators and the doping of which leads to superconductivity. Thus one must solve the problem of "doping the Mott insulator" first. Get the right doing at which AFM is killed. Then one should explain superconductivity systematically building upon this development. 
Then the theory should explain right order of magnitude of $T_c$ and the "dome" shape structure with experimentally determined other required  parameters etc.

\begin{center}
\begin{minipage}{.4\linewidth}
Sample "theoretical minimum" no. 3:  
\end{minipage}
\end{center}

QCP is of central importance! One must first understand the anomalies at this point ($c_{el} \propto -T\ln T$ etc). One must first understand why does the carrier number reduce from $1+p$ holes at $p>p^*$ to $p$ holes at $p<p^*$.  Then one should construct the theory, and then the theory should explain right order of magnitude of $T_c$ and the "dome" shape structure with experimentally determined other parameters etc. 

\begin{center}
\begin{minipage}{.4\linewidth}
etc...
\end{minipage}
\end{center}

Which is the right way to follow? Again, this is a difficult state of affairs! The choice will be affected by one's prejudice and one's vantage point. But the author would like to defend a "minimal theoretical minimum". According to this scheme of "minimal theoretical minimum": (1) a direct experimental verification  (the "smoking gun signature") of the proposed pairing mechanism, and (2) the quantitative reproduction of the value of $T_c$ using experimentally determined other parameters, are sufficient conditions.

And it has been achieved. It is the incontrovertible verdict of the experiment\cite{sea} and settles the pairing problem once and for all. Other issues (related to the strange metal regime) will be resolved with time.

\section{In conclusion}
%%%%%%%%%%%%%%%%%%%%%%%%%%%%%%%%%%%%%%%%%%%%%%%%%%%%%%%%%
The jury is still out on whether quantum criticality motivated approaches have any essential relevance to the key of cuprate superconductivity? In defense, the author would like to argue that as far as the paring mechanism is concerned we have now a direct experimental verification of the charge-transfer superexchange mechanism. So this debate is settled now. It remains to be seen whether in the physics of the strange metal regime, quantum criticality motivated approaches are required or not?

Finally, in this manuscript a brief review of the pivotal experiment is given. The key mechanism of exchange interactions is explained using a very simple model and in a very simple language. We discussed what is settled now and what is not settled yet. We reasoned about a minimal "theoretical minimum" that has been achieved. 

\vspace{3pc}

\section*{Acknowledgments}
Author dedicates this manuscript to the memory of his professor, prof. N. Kumar.

\section*{References}

%----------------------------------------------------------------------------------------
%	REFERENCE LIST
%----------------------------------------------------------------------------------------

%----------------------------------------------------------------------------


\begin{thebibliography}{99} % Bibliography - this is intentionally simple in this template


\bibitem{sea} Shane M. O'Mahony et al, ``On the electron pairing mechanism of copper-oxide high temperature superconductivity", P.N.A.S. {\bf 119}, e2207449119 (2022).


\bibitem{t1}C. Weber, K. Haule, and G. Kotliar, ``Strength of correlations in electron- and hole- doped cuprates", Nat. Phys. {\bf 6}, 574 (2010).


\bibitem{t2} C. Weber, C. Yee, K. Haule, G. Kotliar, ``Scaling of the transition temperature of hole-doped cuprate superconductors with the charge-transfer energy" Europhys. Lett. {\bf 100}, 37001 (2012)


\bibitem{t3} C.-H. Yee, G. Kotliar, ``Tuning the charge-transfer energy in hole-doped cuprates" Phys. Rev. B {\bf 89}, 094517 (2014).


\bibitem{t4} S. Acharyaet al., ``Metal-insulator transition in copper oxides induced by apex displacements" Phys. Rev. X {\bf 8}, 021038 (2018).


\bibitem{t5}Y. Ohta, T. Tohyama, S. Maekawa, ``Electronic structure of insulating cuprates: Role of Madelung potential in the charge-transfer gap and superexchange interaction" Physica C {\bf 185}, 1721-1722(1991).


\bibitem{t6} L. F. Feiner, M. Grilli, C. Di Castro, ``Apical oxygen ions and the electronic structure of the high-$T_c$ cuprates"  Phys. Rev. B {\bf 45}, 10647-10669 (1992).

\bibitem{t7}E. Pavarini, I. Dasgupta, T. Saha-Dasgupta, O. Jepsen, O. K. Andersen, ``Band-structure trend in hole-doped cuprates and correlation with $T_{c, max}$".  Phys. Rev. Lett.  {\bf 87}, 047003 (2001).

\bibitem{t8}K. Foyevtsova, R. Valenti, P. J. Hirschfeld, ``Effect of dopant atoms on local superexchange in cuprate superconductors: A perturbative treatment" Phys. Rev. B {\bf 79}, 144424 (2009).

\bibitem{t9} N. Kowalski etal., ``Oxygen hole content, charge-transfer gap, covalency, and cuprate superconductivity" P.N.A.S. {\bf 118},  e2106476118 (2021).

\bibitem{t10}P. Mai, G. Balduzzi, S. Johnston, T. A. Maier, ``Pairing correlations in the cuprates:  A numerical study of the three-band Hubbard model" Phys. Rev. B {\bf103}, 144514 (2021).


\bibitem{shen}Shengtao Jiang, D. J. Scalapino, and S. R. White, ``Pairing properties of the $t-t^\p-t^{\p\p}-J$ model", arXive:2206.07812v1.


\bibitem{gau}N. Gauquelin etal., ``Atomic scale real-space mapping of holes in $YBa_2Cu_3O_{6+\delta}$", Nat. Comm. 5:4275 (2014).


\bibitem{emery} V. J. Emery, ``Theory of high-$T_c$ superconductivity in oxides"  Phys. Rev. Lett. {\bf 58},2794-2797 (1987).


\bibitem{ander1}. W. Anderson, ``The resonating valence bond state in $La_2CuO_4$ and superconductivity. Science {\bf 235}, 1196-1198 (1987).

\bibitem{ander2}P. W. Anderson, ``Personal history of my engagement with cuprate superconductivity, 1986-2010", Int. J. Mod. Phys. B. {\bf 25}, 1-39 (2011).

\bibitem{zr} F. C. Zhang, T. M. Rice, ``Effective Hamiltonian for the superconducting Cu oxides" Phys. Rev. B {\bf 37}, 3759-3761 (1988).


\bibitem{feyn}R. P. Feynman, ``Lectures on physics", Vol III (Narosa Publishing House, 1986).


\bibitem{pauling}L. Pauling, ``The nature of the chemical bond", Cornell University Press (1939).


\bibitem{ander3}P. W. Anderson etal., ``The physics behind high-temperature superconducting cuprates: the ``plain-vanilla" version of RVB", J. Phys: Cond. Matt. {\bf 16}, R755 (2004).


\bibitem{param}A. Paramekanti, M. Randeria, and N. Trivedi, ``High-$T_c$ superconductors: a variational theory of the superconducting state", Phys. Rev. B. {\bf 70}, 054504 (2004).


\bibitem{zhang}F. C. Zhang, C. Gros, T. M. Rice, H. Shiba, ``A renormalised Hamiltonian approach to a resonant valence bond wavefunction" Supercond. Sci. Technol. {\bf 1},36-46 (1988).


\bibitem{ruck}A. E. Ruckenstein, P. J. Hirschfeld, J. Appel, ``Mean-field theory of high-Tc superconductivity: The superexchange mechanism" Phys. Rev. B {\bf 36}, 857-860 (1987).


\bibitem{kot}G. Kotliar, J. Liu, ``Superexchange mechanism and d-wave superconductivity"  Phys. Rev. B {\bf38}, 5142-5145 (1988).


\bibitem{lee2}P. A. Lee, N. Nagaosa, T.-K. Ng, X.-G. Wen, ``SU(2) formulation of the t-J model: Application to underdoped cuprates" Phys. Rev. B {\bf 57}, 6003-6021 (1998).


\bibitem{scal}R. T. Scalettar, D. J. Scalapino, R. L. Sugar, S. R. White, Antiferromagnetic, charge-transfer, and pairing correlations in the three-band Hubbard model.Phys. Rev. B {\bf 44}, 770-781 (1991).

\bibitem{bas}G. Baskaran, Z. Zou, and P. W. Anderson, ``The resonating valence bond state and high-$T_c$ superconductivity - a mean field theory", Solid State Comm. {\bf 63}, 973 (1987).

\bibitem{bas1}P. W. Anderson, G. Baskaran, Z. Zou, and T. Hsu, ``Resonating-valence-bond theory of phase transitions and superconductivity in $La_2CuO_4-$based compounds", Phys. Rev. Lett. {\bf 58}, 2790 (1987).


\bibitem{lee}P. A. Lee, N. Nagaosa, X.-G. Wen, ``Doping a Mott insulator: physics of high-temperature superconductivity", Rev. Mod. Phys. {\bf 78}, 17 (2006).


\bibitem{anderb}P. W. Anderson, {\it The theory of superconductivity in the high-$T_c$ cuprates}, Princeton University Press (1997).


\bibitem{bat}B. Batlogg, ``Physical properties of high-$T_c$ superconductivity", Phys. Today {\bf 44}, 6 (1991).

\bibitem{bari}N. Barisic etal, ``Universal sheet resistance and revised phase diagram of the cuprate high-temperature superconductors", P.N.A.S. {\bf 110}, 12235 (2013).

\bibitem{taillefer1} L. Taillefer, ``Scattering and pairing in cuprate superconductors", Ann. Rev. Cond. Matt. Phys. {\bf 1}, 50 (2010).


\bibitem{cooper}R. A. Cooper etal, ``Anomalous criticality in the electrical resistivity of $La_{2-x}Sr_xCuO_4$", Science {\bf 323}, 603 (2009).


\bibitem{dago}E. Dagotto, ``Correlated electrons in high temperature superconductors", Rev. Mod. Phys. {\bf 66}, 763 (1994).


\bibitem{ander20}P. W. Anderson, ``Is there glue in cuprate superconductors", Science {\bf 316}, 1705 (2007).


\bibitem{taillefer2}C. Proust and L. Taillefer, ``The remarkable underlying ground states of cuprate superconductors", Ann. Rev. Cond. Matt. Phys. {\bf 10}, 409 (2019).



\bibitem{emery2}V. J. Emery and S. A. Kivelson, ``Importance of phase fluctuations in superconductors with small superfluid density", Nature {\bf 374}, 434 (1995).



\bibitem{nav}Navinder Singh, ``Leading theories of the cuprate superconductivity: A critique", Physica C: Supercond. and its Appls. {\bf 580}, 1353782 (2021).


\bibitem{mflp}C.M. Varma etal, ``Phenomenology of the normal state of Cu-O high-temperature superconductors",  Phys. Rev. Lett. {\bf 63} 1996 (1989).




\end{thebibliography}
\end{document}